# Determining the value of the fine-structure constant from a current balance: getting acquainted with some upcoming changes to the SI

Richard S. Davis*

*International Bureau of Weights and Measures (BIPM), Sèvres, France 92312*

Abstract

The revised International System of Units (SI), expected to be approved late in 2018, has implications for physics pedagogy. The ampere definition which dates from 1948 will be replaced by a definition that fixes the numerical value of the elementary charge, *e*, in coulombs. The kilogram definition which dates from 1889 will be replaced by a definition that fixes the numerical value of the Planck constant, *h*, in joule seconds. Existing SI equations will be completely unaffected. However, there will be a largely-negligible, but nevertheless necessary, change to published numerical factors relating SI electrical units to their corresponding units in the Gaussian and other CGS systems of units. The implications of the revised SI for electrical metrology are neatly illustrated by considering the interpretation of results obtained from a current balance in the present SI and in the revised SI.

## I. INTRODUCTION

What is now known as the International System of Units (SI) adopted the ampere more than 60 years ago, preceded by lengthy discussions and a recommendation from the International Electrotechnical Commission (IEC).[1,2] The definition of the ampere famously relies on the specification of a force per unit length of wire of a current balance that is manifestly impossible to construct. However, a simple current balance found in many undergraduate physics laboratories can be used to illustrate the ampere definition by measuring electric current and confirming that the fixed numerical value of the magnetic permeability of free space expressed in $N/A^2$ leads to a result that agrees with that of a calibrated ammeter.[3]

Four of the seven base units of the SI are expected to be redefined in about two years' time: the kilogram, ampere, kelvin and mole to create a system that is "more fundamental."[4,5] The first two of the new definitions will change how measurements with a current balance are interpreted, and these will be the focus of this article. The definitions of the remaining three base units of the present SI (the second, the meter, and the candela) will not be revised. The big picture is given in Refs. 4 and 5. The following analysis shows that the ideal current balance invoked in the present definition of the ampere will in the near future yield an experimental value for the fine-structure constant instead.

In the following, the term "present" SI refers to the system of units defined in Ref. 1, which are in general use; the term "revised" SI refers to the revisions expected to be approved in the autumn of 2018.[4,5] After its implementation, the revised SI will simply be "the SI."

## II. WHAT BECOMES OF THE CURRENT BALANCE IN THE REVISED SI?

Ever since the SI was introduced, the ampere has been defined in terms of the measurement of a mechanical force. A version of the current balance can be found in many undergraduate physics laboratories and is used to measure an electric current in SI units, demonstrating how the present ampere definition can be realized.[3,6,7] In the revised SI, the unit of charge, the coulomb, will be defined by giving a fixed (i.e. exact) numerical value to the elementary charge, $e$, when expressed in coulombs (by convention, the electron has charge $-e$).[4,5]

The idealized current balance with infinitely long parallel wires, whose description appears in the present definition of the ampere, will no longer be mentioned. Rather, in the revised SI, the ampere will be realized from Ohm's law by measuring the voltage $U$ across a resistance $R$. The voltage will be measured in terms of the ac Josephson effect:[4,5,8]

$$U = n\frac{h\nu}{2e} = n\frac{\nu}{K_J}, \qquad (1)$$

where $h$ is the Planck constant, $\nu$ is an adjustable microwave frequency, and $n$ is an integral number of voltage steps. The Josephson constant, $K_J$, has the SI unit Hz/V. The resistance is measured using the quantum-Hall effect:[4,5,9,10]

$$R = \frac{1}{i}\frac{h}{e^2} = \frac{1}{i}R_K, \qquad (2)$$

where $i$ is an integral number of impedance steps and $R_K$, whose SI unit is $\Omega$, is known as the von Klitzing constant. The constants $h$ and $e$ appear in both equations.

From Eqs. (1) and (2), the measured current $I$ is

$$I = \frac{U}{R} = \frac{in}{2}e\nu. \qquad (3)$$

Since 1967, SI frequency measurements have been traceable to $\Delta\nu_{Cs}$, the hyperfine transition frequency of the cesium atom, whose numerical value in hertz has been fixed to define the SI second.[1] In the revised SI, the numerical values of $h$ and $e$ will also be fixed to define the SI units J s (kg m² s⁻¹) and C (A s) respectively. Assigning a fixed numerical value to $h$ also has the effect of redefining the kilogram because the second and meter are already defined.[4] (In 1983 the numerical value of $c$, the speed of light in vacuum, was given a fixed numerical value in m/s to define the meter.[11]) The revised definition of the kilogram will no longer depend on the mass of a unique object manufactured in the 19th century, the international prototype of the kilogram (IPK).[1,12] Continuity between the present and future definitions of the kilogram will be established by ensuring that the numerical value of $h$ fixed in the revised SI equals the best experimental value of $h$ in the present SI.[13] The continuity between present and future definitions of the ampere is being established by the same strategy, as shown in detail below, using the current balance as an illustration.

If the current balance will no longer exemplify the definition of the ampere as it does at present, what becomes of this metrological device and its simplified version used for laboratory demonstrations? We now answer this question.

We begin with the form of Ampère's force law used to define the ampere at present,

$$F_L = 2k_A \frac{I^2}{a}, \qquad (4)$$

where $F_L$ is the force per unit length between two infinitely long parallel wires of negligible cross section, separated by a distance $a$ and maintained in vacuum.[1,14] The constant $k_A$ is specific to the system of units that has been adopted, whereas the factor 2 is a geometric term applicable to the special case of the magnetic field produced by an infinitely long wire with current $I$ flowing in it. Equation 4 shows the Lorentz force on a second wire, parallel to the first and carrying the same current.[15] The wires attract or repel depending on whether current in them flows in the same or opposite direction. The ampere definition specifies that if $a = 1$ m, a current of 1 A flowing in each wire will result in $F_L$ having a magnitude of $2 \times 10^{-7}$ N/m. In the SI, $k_A = \mu_0/4\pi$. The present definition of the ampere is therefore equivalent to the specification $\mu_0/4\pi \equiv 10^{-7}$ N/A$^2 \equiv 10^{-7}$ H/m. The constant $\mu_0$ is usually referred to as the permeability of free space. These choices are the same as those of the rationalized MKS system of units, also known as the MKSA system, which predates the SI.[1] In the Gaussian-CGS system, $k_A = 1/c^2$.[14] Therefore electrical current in the Gaussian system has the unit dyn$^{1/2}$ cm s$^{-1}$. The importance of Eq. (4) in what follows is its role in defining the ampere in the present SI, whereas the ampere will be defined differently in the revised SI. Although a continuity condition will ensure that the magnitude of the ampere will be essentially unchanged by the revision, there will be interesting consequences.

In a typical device used to illustrate the Ampere force equation,[3] the length $L$ of each wire is finite but still reasonably long compared with $a$, i.e. $L \gg a$. In that case the SI version of Eq. (4) can be approximated by

$$\frac{F}{L} \cong 2\left(\frac{\mu_0}{4\pi}\right)\frac{I^2}{a} \tag{5}$$

where $F$ is a force determined by the weight $mg$ of a mass $m$ counter-acting the force between the wires, where $g$ is the gravitational acceleration in the laboratory. Given that the relative uncertainty of $I$ determined from these devices is typically of order 1 %, there is no need to introduce the correction of less than one part in $10^6$ for the permeability of air relative to that of vacuum. Since the definition of the ampere agreed in 1948, $\mu_0/4\pi$ has had the exact value $10^{-7}$ N/A$^2$. In the revised SI, the factor $10^{-7}$ can no longer be exact because this would conflict with the decision to fix the numerical value in coulombs of the elementary charge $e$. When the SI coulomb is redefined in this way, the ampere is also redefined since 1 C = 1 A s. The numerical value of $\mu_0/4\pi$ in N/A$^2$ must then be determined by experiment. Nevertheless, the value of $e$ chosen will be consistent with $\mu_0/4\pi = 10^{-7}$ N/A$^2$ to within the smallest possible uncertainty at the time the revised SI is adopted. In fact, the relative uncertainty of $\mu_0$ will be identical to the relative uncertainty of the fine-structure constant, $\alpha$, which is a few parts in $10^{10}$ in the most recent recommendation.[13]

How did the fine-structure constant come into this picture? We can see by examining Eq. (4) or (5). We start with the right-hand side, which is the same for both equations. In the SI, present and revised, $a = b_1(c/\Delta\nu_{Cs})$, where $b_1$ is a pure number (also referred to in this context as a dimensionless number or a number of dimension one);[1] $c$ and $\Delta\nu_{Cs}$ have fixed values in SI units, as shown in Tables I and II. Thus an exact specification of $a$ in meters is equivalent to an exact specification of $b_1$.

In the revised SI, $I = b_2(e\Delta\nu_{Cs})$, where $b_2$ is a pure number and $e$ will have a fixed value. The recommended value for $e$,[13] shown in Table I, is taken to be exact for demonstration purposes.

We also use the following relation among physical constants, discussed in Section 3, which is valid today and must remain valid in the revised SI:

$$\frac{\mu_0}{4\pi} = \frac{\alpha}{2\pi}\frac{h}{e^2 c}. \tag{6}$$

The constants of physics are quantities; the relations between physical quantities are the equations of physics, which reflect our scientific understanding. By contrast, definitions of units are generally decided by committees. From Eq. (6), we see again why a committee of SI experts would not modernize the unit system by assigning fixed numerical values to $c$, $h$, $e$, <u>and</u> $\mu_0$. Not only would this define the ampere in two different ways but the fine-structure constant, which must be the same experimentally-determined dimensionless number in every system of units, would acquire a fixed value in the revised SI. Clearly $\mu_0$ must become an experimentally determined quantity. But we can also see from Eq. (6) that a measurement of $\mu_0$ will be equivalent to a measurement of $\alpha$ (and vice versa) when $h$, $e$ and $c$ all have fixed values.

It is nevertheless instructive to make the substitutions defined above in Eq. (4). For the right-hand side,

$$F_L = \frac{\alpha}{\pi} \frac{b_2^2}{b_1} \frac{h(\Delta \nu_{Cs})^3}{c^2}. \tag{7}$$

Note that the charge $e$ has dropped out of the result, but that its value was used to determine the number $b_2$, which remains. This would also be true for Eq. (5).

Now look at the left-hand side of equation Eq. (4) (the same result would also be obtained from Eq. (5)). The ratio of force to length in the SI (present and revised) has the unit N/m = kg s$^{-2}$. Since 1889, the kilogram has been defined in terms of the mass $m$(IPK) of the international prototype of the kilogram (IPK), a unique object made of a platinum alloy;[12] in the revised SI, however, $F_L = b_3 \left( h(\Delta \nu_{Cs})^3 / c^2 \right)$, where $b_3$ is a pure number. To calculate $b_3$ we again refer to the 2014 CODATA recommendation[13] for the value of $h$ and pretend that this value has already been fixed to have an exact value (no uncertainty). Then Eq. (7) becomes

$$b_3 \frac{h(\Delta \nu_{Cs})^3}{c^2} = \frac{\alpha}{\pi} \frac{b_2^2}{b_1} \frac{h(\Delta \nu_{Cs})^3}{c^2} \tag{8}$$

where all dimensioned quantities have fixed values that define the units of interest to this discussion in the revised SI, including $e$ whose fixed value was needed to calculate $b_2$. Finally,

$$\alpha = \pi \frac{b_1 b_3}{b_2^2}. \tag{9}$$

Table II shows the calculation of the fine-structure constant from Eq. (9) for the parameters $a$, $I$ and $F_L$ specified in the present definition of the ampere, which assumes that the parameters are known without any error. The result is indeed the present value of the fine-structure constant according to the latest recommendation of the CODATA Task Group on Fundamental Constants.[13] The last row of the table uses the CODATA convention that (17) is the standard uncertainty of 64, the last two digits shown. That the deduced and published values of $\alpha$ are not identical is an indication that the present numerical values of $h$ and $e$ do have an associated experimental uncertainty although we have pretended that they do not. In an idealized experiment carried out in the revised SI, if the values of $I$ and $a$ are specified without error, $F_L$ must nevertheless be measured in order to determine $\mu_0$ or $\alpha$. At the time the revised SI takes effect, continuity conditions will lead to the result calculated in Table II within the measurement uncertainty of $\alpha$.

The designation of a certain set of units as SI "base units"[1,4] and all other units as "derived" will lose some of its importance in the revised SI.[5] Of primary importance will be the seven defining constants,[4,5] although only the subset $\Delta\nu_{Cs}$, $c$, $h$, and $e$ shown in Table I has been needed to discuss the current balance. The numbers $b_1$ and $b_2$ are perfectly consistent with the definitions of the meter in the present and revised SI and the ampere in the revised SI, e.g. $a = 1$ m is equivalent to $a = b_1(c/\Delta\nu_{Cs})$. In this case, $b_1$ is the fixed numerical value of $\Delta\nu_{Cs}/c$ stripped of its SI unit (m$^{-1}$), which we may write as $\{\Delta\nu_{Cs}/c\}$. There is no conceptual difference between the specifications for 1 m and 1 A, and that of $2 \times 10^{-7}$ N/m, even though the N/m is not one of the seven base units of the SI. They all rely on the defining constants.

Current balances used by students demonstrate the definition of the ampere in the present SI. Students can verify that the ampere calculated from the fixed value of $\mu_0$ agrees with the current measured using a calibrated ammeter. The apparatus can be used to measure $\mu_0$ in the revised SI by assuming that the calibration of the ammeter is traceable to quantum electrical standards, thereby showing that the value of $\mu_0$ derived from the current balance is consistent with the recommended value. Using measurement units as defined in the revised SI, the current balance also measures $\alpha$ with the same relative uncertainty as $\mu_0$ via Eq. (6). Although the different interpretations of the results obtained in the present and revised SI follow from real conceptual distinctions, observed differences between the present and revised SI will be invisible at the uncertainties attainable by demonstration equipment. This is indeed as it should be since great care is being taken to ensure that such differences will pass undetected by the vast majority of scientists and technicians, and by all of the general public. Nevertheless, the advantages of modernizing the SI are real, as discussed in Refs. 4, 5 and 16.

### III. THE FINE-STRUCTURE CONSTANT IN OTHER UNIT SYSTEMS

The fine-structure constant first emerged from Bohr's model of the hydrogen atom, from which one may deduce

$$\alpha = k_C \left( e^2 / \hbar c \right) \quad (10)$$

where $\hbar = h/2\pi$ and $k_C$ depends on which equations of electrostatics were used to model the force between the proton and electron (see Appendix). In the Gaussian CGS system, almost always preferred by theoretical physicists over the SI,[14,15,17] the fine-structure constant is given by

$$\alpha = e'^2 / \hbar c \quad \text{(Gaussian)} \quad (11)$$

where $e'$ is the elementary charge in statcoulomb (statC), $\hbar$ is in erg seconds (erg s) and $c$ is the speed of light in vacuum in centimeters per second (cm/s). Equation (11) also applies to the CGS-esu system. The corresponding SI equation is

$$\alpha = \left(1/4\pi\varepsilon_0\right)\left(e^2/\hbar c\right) = \left(\mu_0 c/2\right)\left(e^2/h\right) \quad \text{(SI)} \tag{12}$$

where $e$ is in coulombs (C), $h$ is in joule seconds (J s) and $c$ is in meters per second (m s$^{-1}$).[13] Equation (6) is just a rearrangement of Eq. (12). We use the symbol $e'$ for the Gaussian value of the elementary charge because, while the conversion factors 1 kg = $10^3$ g and 1 m = $10^2$ cm are obvious, the conversion factors for electromagnetic units are less so, as discussed in Ref. 14 and the Appendix.

Thus, in equations adopted by the SI, $k_C = (4\pi\varepsilon_0)^{-1}$ where $\varepsilon_0$ is the permittivity of free space. Its unit is the farad per meter, F/m; $k_C = 1$ in the Gaussian system. The constraint $k_C / k_A = c^2$ applies to both systems.[14] We may always use the substitution $\mu_0 c^2 = 1/\varepsilon_0$ in SI equations such as Eq. (12). Once again, no matter which unit system we use (and there are many other possibilities not presented here), $\alpha$ as defined by Eq. (10) is always the same experimentally-determined, dimensionless number.

As shown in the Appendix, in the present SI the charge correspondence between the systems of units is: $10\{c\}_{SI}\, \text{statC} \leftrightarrow 1\, \text{C}$, where curly brackets are used to indicate that $\{c\}_{SI}$ is only the numerical value of the speed of light in vacuum when expressed in m/s. At present, this conversion factor is exact. In the revised SI the value of $\mu_0 / 4\pi$ will still be $10^{-7}$ N/A$^2$ at the time of adoption (see Table II), but subsequently the value of $\mu_0 / \alpha$ will be fixed. The recommended value and uncertainty of $\mu_0$ must then evolve in perfect correlation with improved experimental determinations of $\alpha$. The most accurate determinations of $\alpha$ will not come from current balances but rather will continue as in recent years to be supplied by experiments involving atomic and subatomic physics.[18]

## IV. WHY THE APPEARANCE OF $\alpha$ IN THE ANALYSIS OF THE CURRENT BALANCE IS NOT DUE TO NEW PHYSICS

Measurement units are derived from the accepted equations of physics, and these impose certain logical constraints. Once these constraints are satisfied, remaining choices for a unit system can (and must) be arbitrary, to paraphrase J. D. Jackson.[14] The SI has aimed to satisfy a broad spectrum of physicists, engineers and chemists, while still accommodating the needs of the general public. To continue with Jackson, "The desirable features of a system of units in any field are convenience and clarity." His own choices evolved. By the 3rd Edition of *Classical Electrodynamics*, he continued to use the Gaussian system beginning with Chapter 11, whereas in the preceding chapters he opted for the SI, advising physicists to be conversant with both.

The only reason a device like the current balance can be used to determine a value for $\alpha$ in the revised SI is the choice of the fixed constants used to define the system of units. All measurements made with the revised SI will be traceable to these fixed constants. The arbitrary choices have respected the logical constraints, one of the latter being that the fine-structure constant cannot have a fixed value.

Revising the definitions of two measurement units (kg and A) in the manner described in Section II has of course been motivated by the discovery of physical phenomena. One of these is the quantum-Hall effect. Due to the widespread use in electrical metrology of this effect, Eq. (12) is sometimes written as

$$\alpha = \frac{\mu_0 c}{2R_K} \tag{13}$$

where the quantity $R_K$ is now routinely accessible in many laboratories. In their initial publication, von Klitzing and co-authors stressed that Eq. (13) pointed the way to a novel and accurate (for 1980) method to determine the value of $\alpha$ by measuring $R_K$ in terms of the present realization of the SI ohm.[19] They also pointed out in the same publication that, should $\alpha$ be known to smaller uncertainty from a different experiment, then the quantum-Hall effect might still be used to derive a known standard resistance. From Eq. (13) we see that in the present SI the quantity $R_K \alpha$ has a fixed numerical value. In the revised SI, the numerical value of $R_K$ itself will be fixed, leading to a definition of the ohm. This definition can be realized via Eq. (2). In fact, the volt will also be defined because the numerical value of $h\Delta v_{Cs}/e$ will be fixed (Eq. (1)). The definition of the ampere follows from the fixed numerical value of $e\Delta v_{Cs}$, which can be realized via Eq. (3) or any other suitable method.[20] Another feature of Eq. (12) is that, in the SI, $\mu_0 c = \sqrt{\mu_0/\varepsilon_0} = Z_0$ or

$$\alpha = \frac{Z_0}{2R_K}. \tag{14}$$

The constant $Z_0$ is also identified in the SI as "the characteristic impedance" of free space, about 377 Ω (see Ref. 14, p. 297). The corresponding electromagnetic "wave impedance" in free space is the number 1 in the Gaussian system.[21] Nevertheless, Eq. (14) applies to both the SI and the Gaussian system but the relation $Z_0 = 4\pi/c$ must be used in the latter, as can be inferred from Eq. (11). This leads to the correspondence $\{\mu_0 c\}_{SI}\,\Omega \leftrightarrow \{4\pi/c\}_{Gauss}$ s/cm from which the conversion factor[14,15] between a numerical value of resistance in Ω and its corresponding numerical value in s/cm, the Gaussian unit of resistance, is derived.

## V. SUMMARY

In the present SI, the definitions of the second, meter, kilogram and ampere are equivalent to specifying fixed numerical values for the physical constants $\Delta\nu_{Cs}$ in Hz, $c$ in m/s and $\mu_0$ in N/A$^2$, as well as specifying the mass of the international prototype of the kilogram, $m$(IPK), to be 1 kg exactly. Various combinations of these quantities are used to define the second, meter, kilogram, and ampere. In this system, an idealized current balance has served to define the ampere, with the fixed value of $\mu_0$ implicit in the ampere definition. A student version of the current balance is often used to illustrate the principles involved. One might have said that once the units of mass, length and time have been defined, the further specification of an exact numerical value for $\mu_0$ in the unit N/A$^2$ is sufficient to define the ampere, with analysis confirming that the current balance is one way to realize this definition.

A revised SI is expected to be approved in the autumn of 2018 and to take effect relatively soon thereafter. In this system, the fixed value of $\mu_0$, $4\pi \times 10^{-7}$ N/A$^2$, is replaced by a fixed numerical value of $e$ thereby defining the coulomb; the fixed value of $m$(IPK), 1 kg, is replaced by a fixed numerical value of $h$ thereby defining the joule second. The numerical value of $\Delta\nu_{Cs}$ has been fixed since 1967 to define the second and the numerical value of $c$ has been fixed since 1983 to define the meter per second. The quantities $c$, $e$ and $h$ appear in the defining relation of the fine-structure constant, $\alpha$, which raises some interesting questions about the nature of units and the role of physical constants such as $\alpha$, whose value must be independent of any unit system. Some of the issues involved have been presented by asking the question "what does a current balance measure?" We also show an instance of how the fine-structure constant influences the revised SI: At present, $\mu_0$ and $R_K\alpha$ have exact values; in the revised SI, the values of $R_K$ and $\mu_0/\alpha$ will be exact. Conversion factors to CGS systems which presently make use of the exact relation $\{\mu_0/4\pi\} \equiv 10^{-7}$ will no longer be strictly correct after the revised SI takes effect.

**Appendix:** Conversion between statcoulomb and coulomb in the present and revised SI

First we derive the well-known conversion factor between statcoulomb (statC) in the Gaussian system, and the coulomb (C) in the present SI. Take the example of two point charges, each with charge $q = 1$ C and separated by 1 m. The force between the charges, about 9 GN (!), is given by

$$F = \left\{k_C \frac{q^2}{r^2}\right\}_{SI} N = \left\{\frac{1}{4\pi\varepsilon_0}\right\}_{SI} N = \left\{\frac{\mu_0}{4\pi}c^2\right\}_{SI} N. \tag{A1}$$

The curly brackets signify that we extract only the numerical values of the quantities within. This must be the same force measured in dyn (CGS), where the Gaussian charge is $q'$, the separation $r$ is $10^2$ cm and $k_C = 1$. Thus, $F = \{q'^2/10^4\}_{Gauss}$ dyn $= \{\mu_0 c^2/4\pi\}_{SI}$ N. Since 1 N = $10^5$ dyn,

$$\{q'\} \text{statC} = 10^4 \left\{10\frac{\mu_0}{4\pi}c^2\right\}_{SI}^{\frac{1}{2}} \text{statC} \leftrightarrow 1\text{ C}. \tag{A2}$$

Since $\{\mu_0/4\pi\} \equiv 10^{-7}$ in the present SI, the correspondence

$$10\{c\}_{SI} \text{statC} = 2\,997\,924\,580 \text{ statC} \leftrightarrow 1\text{ C} \tag{A3}$$

is a compact and exact form found in conversion tables. In the revised SI, Eq. (A2) will remain exact but $\{\mu_0/4\pi\}$ will become an experimental value with a finite but largely negligible uncertainty. Thanks to the continuity conditions that will be imposed, the experimental value will still be $10^{-7}$ but with a relative uncertainty of a few parts in $10^{10}$.

---

[11] Prior to 1983, when the meter per second was defined independently of $c$, the speed of light could be measured. The venerable history of such measurements ended in 1983. Similarly, a measurement of $\Delta\nu_{Cs}$ in hertz became a logical impossibility as of 1967. It will also become meaningless to measure $h$ and $e$ in SI units (as well as the Boltzmann and Avogadro constants, which do not concern us here) after the revised SI is implemented.

[12] Richard S. Davis, Pauline Barat, and Michael Stock, "A brief history of the unit of mass: continuity of successive definitions of the kilogram," Metrologia, **53** (5), A12-A18 (2016).

[13] Peter J. Mohr, David B. Newell, and Barry N. Taylor, "CODATA recommended values of the fundamental constants: 2014," Rev. Mod. Phys., **88** (3), 035009 (2016). A recommendation appropriate to the revised SI will be generated prior to the formal approval of the revisions.

[14] John David Jackson, *Classical Electrodynamics,* Third Edition (John Wiley & Sons, 1999), Appendix, p. 775-784. In this paper, we modify Jackson's notation so that his $k_1$ is our $k_C$ and his $k_2$ is our $k_A$. Our subscripts serve as mnemonics for Coulomb and Ampère.

[15] David. J. Griffiths, *Introduction to Electrodynamics*, Fourth Edition (Pearson, 2013), Example 5.5, p. 225-226. Conversion factors between SI and Gaussian units are found in Appendix C of this textbook, as well as in Ref. 14.

[16] Nick Fletcher, Gert Rietveld, James Olthoff, Ilya Budovsky, and Martin Milton, "Electrical units in the New SI: Saying goodbye to the 1990 values," NCSLI Measure J. Meas. Sci., **9** (3), 31-35 (2014).

[17] See, for instance, a pithy dismissal of SI conventions in: Frank Wilczek, "On absolute units, II: Challenges and responses," Physics Today, **59** (1), 10-11 (2006).

TABLE I. Four of seven fixed constants in revised SI

| Constants | Their fixed values | |
|---|---|---|
| $\Delta\nu_{Cs}$ | 9 192 631 770 Hz | (a) |
| $c$ | 299 792 458 m s$^{-1}$ | (a) |
| $h$ | 6.626 070 040 × 10$^{-34}$ J s | (b) |
| $e$ | 1.602 176 6208 × 10$^{-19}$ C | (c) |

[a] No change from present SI; used to define the second and the meter.

[b] Value taken from Ref. 13 for demonstration; a fixed value of $h$ will replace the identity $m(\text{IPK}) \equiv 1$ kg, used at present to define the kilogram.

[c] Value taken from Ref. 13, for demonstration; a fixed value of $e$ will replace the identity $\mu_0/4\pi \equiv 10^{-7}$ N/A$^2$, used at present to define the ampere.

TABLE II. Analysis of Ampere balance in revised SI

| Constants | Their fixed values |
|---|---|
| $c/\Delta\nu_{Cs}$ | 0.032 612 255 717 4941 m |
| $e\Delta\nu_{Cs}$ | 1.472 821 970 551 73 × 10$^{-9}$ A |
| $h(\Delta\nu_{Cs})^3/c^2$ | 5.727 092 637 972 45 × 10$^{-21}$ N/m |

| Parameters | Their assumed values |
|---|---|
| $a$ | 1 m |
| $I$ | 1 A |
| $F_L$ | 2 × 10$^{-7}$ N/m |
| $b_1$ | 30.663 318 988 4984 |
| $b_2$ | 6.789 686 873 189 37 × 10$^8$ |
| $b_3$ | 3.492 173 300 531 87 × 10$^{13}$ |

| $\alpha$, deduced from continuity conditions versus published value | |
|---|---|
| $\pi b_1 b_2^{-2} b_3$ (deduced) | 7.297 352 5662 × 10$^{-3}$ |
| $\alpha$ (Ref. 13) | 7.297 352 5664(17) × 10$^{-3}$ |